\documentclass[conference,10pt]{IEEEtran}
\IEEEoverridecommandlockouts

\usepackage{textcomp}
\usepackage{gensymb}
\usepackage{multirow}
\usepackage[ruled,linesnumbered,noend]{algorithm2e} 

\usepackage{cite}
\usepackage{subcaption}
\usepackage{amsmath,amssymb,amsfonts}
\usepackage[ruled,linesnumbered,noend]{algorithm2e} 
\usepackage{graphicx}
\usepackage{textcomp}

\usepackage{dblfloatfix}

\usepackage{tabularx}

\usepackage{url}
\def\BibTeX{{\rm B\kern-.05em{\sc i\kern-.025em b}\kern-.08em
    T\kern-.1667em\lower.7ex\hbox{E}\kern-.125emX}}

\begin{document}

\title{ Parameter Modeling for Small-Scale Mobility in Indoor THz Communication \\
\thanks{This project was funded by CMU Portugal Program: CMU/TMP/0013/2017- THz Communication for Beyond 5G Ultra-fast Networks. We acknowledge Dr. Kazi Mohammed Saidul Huq (from Instituto de Telecomunicações, Portugal) in his early work in discussing this concept with the authors}
}

\author{\IEEEauthorblockN{Rohit Singh \textsuperscript{1}, Douglas Sicker \textsuperscript{1,2}}
\IEEEauthorblockA{ \textsuperscript{1}  \textit{Engineering \& Public Policy}, \textit{Carnegie Mellon University}, Pittsburgh, USA \\
			       \textsuperscript{2}  \textit{School of Computer Science}, \textit{Carnegie Mellon University}, Pittsburgh, USA \\
Email: rohits1@andrew.cmu.edu, sicker@cmu.edu}
}

\maketitle

\section*{ABSTRACT}

Despite such challenges as high path loss and equipment cost, THz communication is becoming one of the potentially viable means through which ultra-high data rate can be achieved. To compensate for the high path loss, we present parameter modeling for indoor THz communication. To maximize efficient and opportunistic use of resources, we analyze the potential workarounds for a single access point to satisfy most of the mobile terminals by varying such parameters as humidity, distance, frequency windows, beamwidths, antenna placement, and user mobility type. One promising parameter is antenna beamwidth, where narrower beams results in higher antenna gain. However, this can lead to ``\textit{beamwidth dilemma}" scenario, where narrower beamwidth can result in significant outages due to device mobility and orientation. In this paper, we address this challenge by presenting a mobility model that performs an extensive analysis of different human mobility scenarios, where each scenario has different data rate demands and movement patterns. We observe that for mobile users, there are optimal beamwidths that are affected by the mobility type (high mobility, constrained mobility, and low mobility) and AP placement.

\section*{Keywords}  Terahertz, Indoor Model, Adaptive Frequency Window, Antenna Beamwidth, Beamwidth Dilemma, AP Placement.

\section{Introduction}
In the near future, the current 4G, WiFi systems, and even 5G systems will not be able to meet the anticipated demand for higher throughput and Quality of Service (QoS) \cite{6GFront}. Even with the best signal-to-interference-noise ratio (SINR), there is a theoretical limit to the achievable data rate, which is directly proportional to the available bandwidth. Obtaining additional bandwidth is one of the primary ways to achieve higher throughput. The mmWave ($10-300 GHz$) bands are already being explored as a means to achieve a system average of $3-7 Gbps$ data rate \cite{mmWaveTest}. The next-generation IEEE 802.11ay is focused explicitly on catering to applications, such as VR and AR games \cite{80211ay}. A more promising solution is to move to terahertz (THz) band ($300 GHz -10THz$) that can help provide higher throughput, in myriad number of applications including recreational indoor applications (e.g., gaming), educational purposes in schools, business communications, hologram communication, 8KTV, hospital patient monitoring, and public safety communications \cite{TeraNetIan}.

\setlength\belowcaptionskip{-0.2 in}
\setlength{\abovecaptionskip}{0.2 in}
\begin{figure*}[t]
\centering
\begin{subfigure}[]{1.6 in}
\includegraphics[width=1.6 in,height=1.7 in]{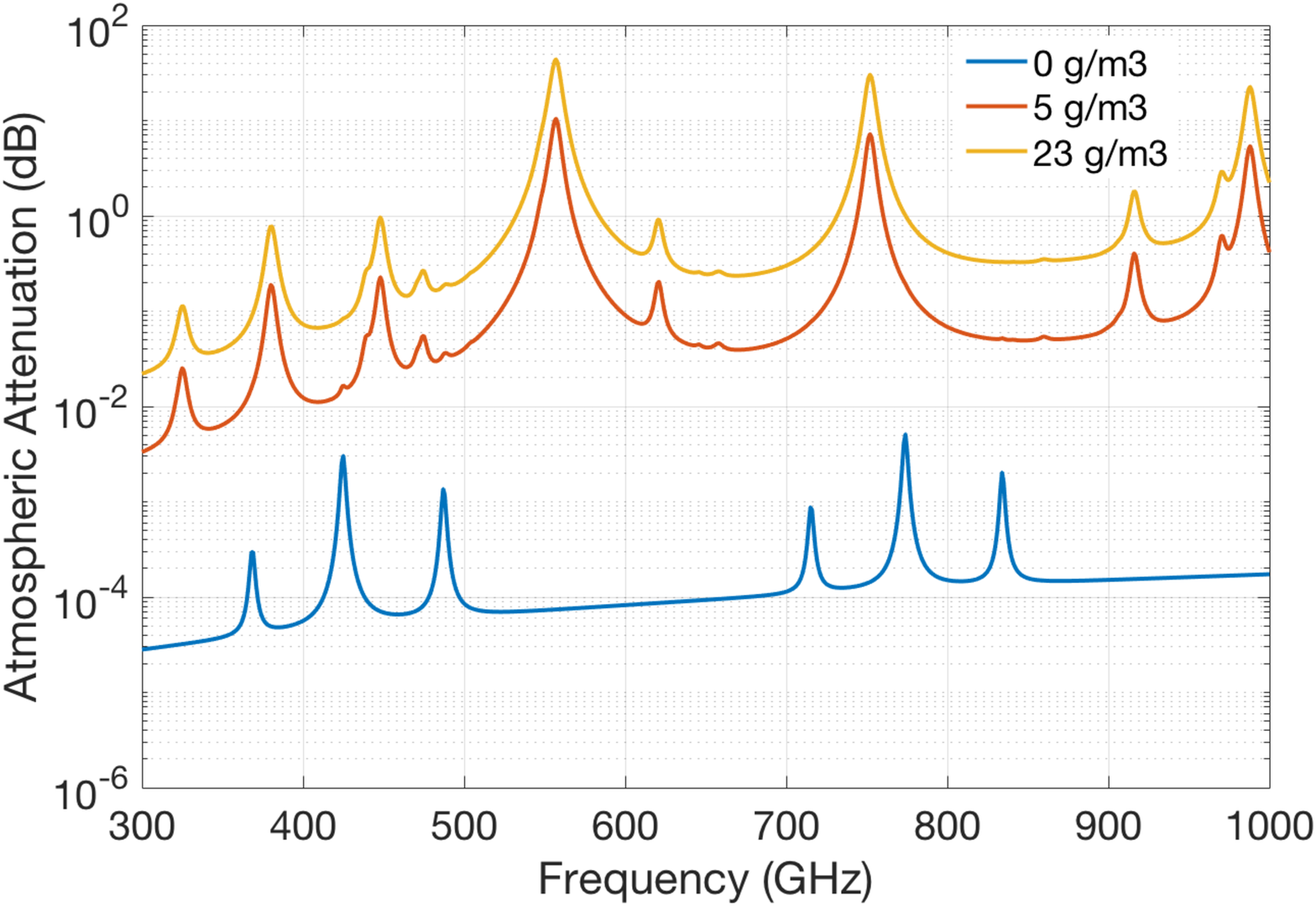}
\caption{$L_A$ at $d_j=1m$}
\label{La1m}
\end{subfigure}
~
\begin{subfigure}[]{1.6 in}
\includegraphics[width=1.6 in,height=1.7 in]{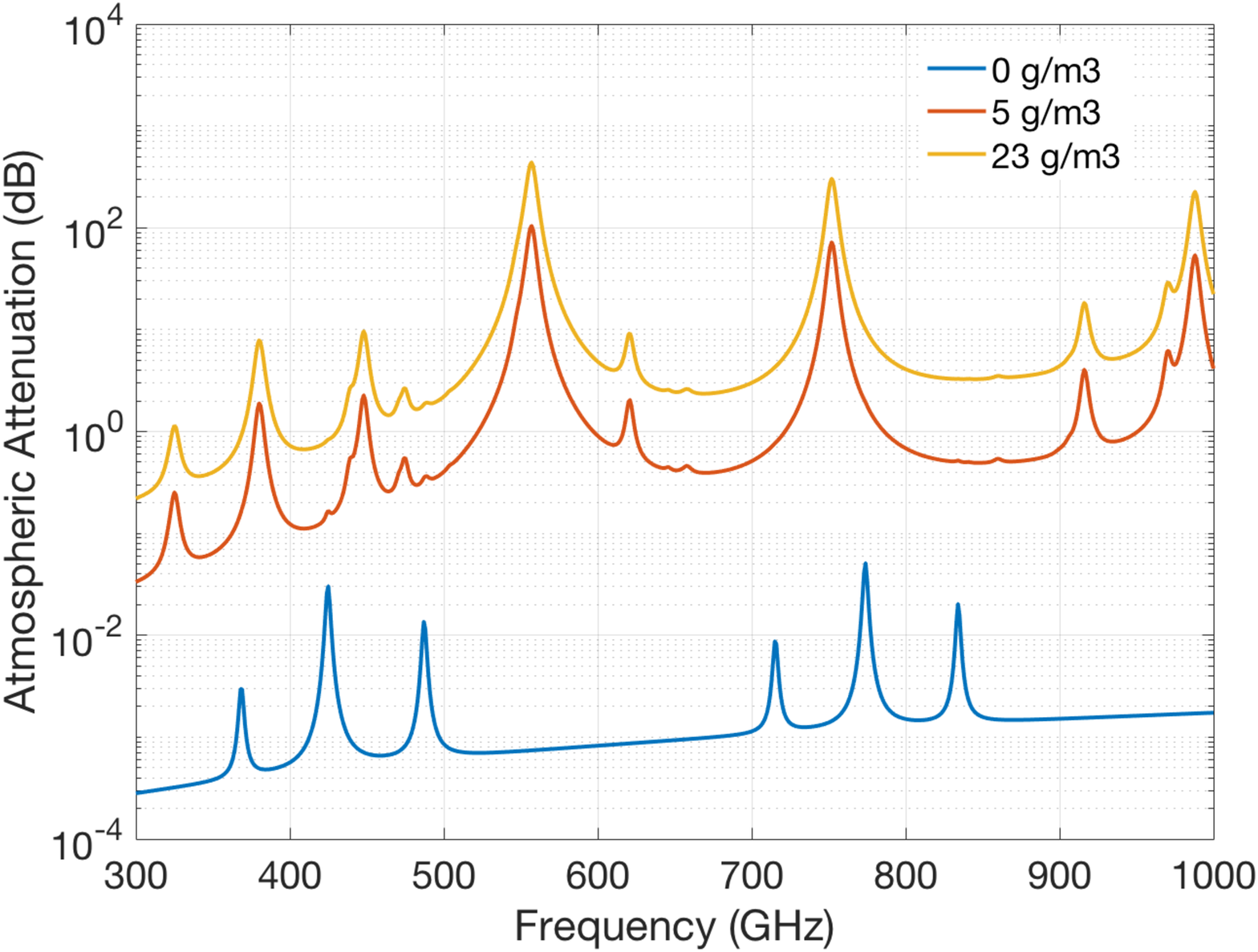}
\caption{$L_A$ at $d_j=10m$}
\label{La10m}
\end{subfigure}
~
\begin{subfigure}[]{1.6 in}
\includegraphics[width=1.6 in,height=1.7 in]{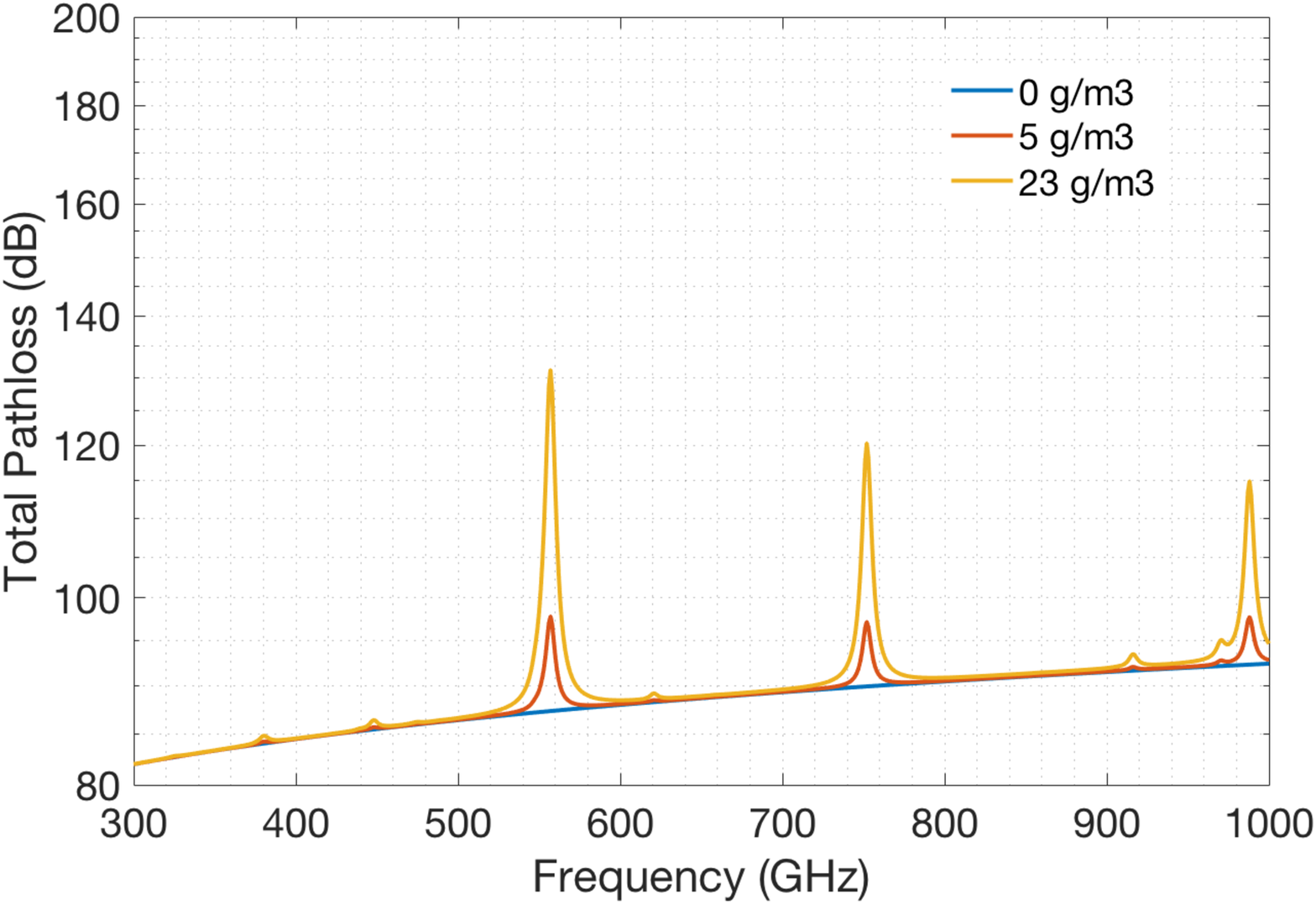}
\caption{$L_T$ at $d_j=1m$}
\label{Lt1m}
\end{subfigure}
~
\begin{subfigure}[]{1.6 in}
\includegraphics[width=1.6 in,height=1.7 in]{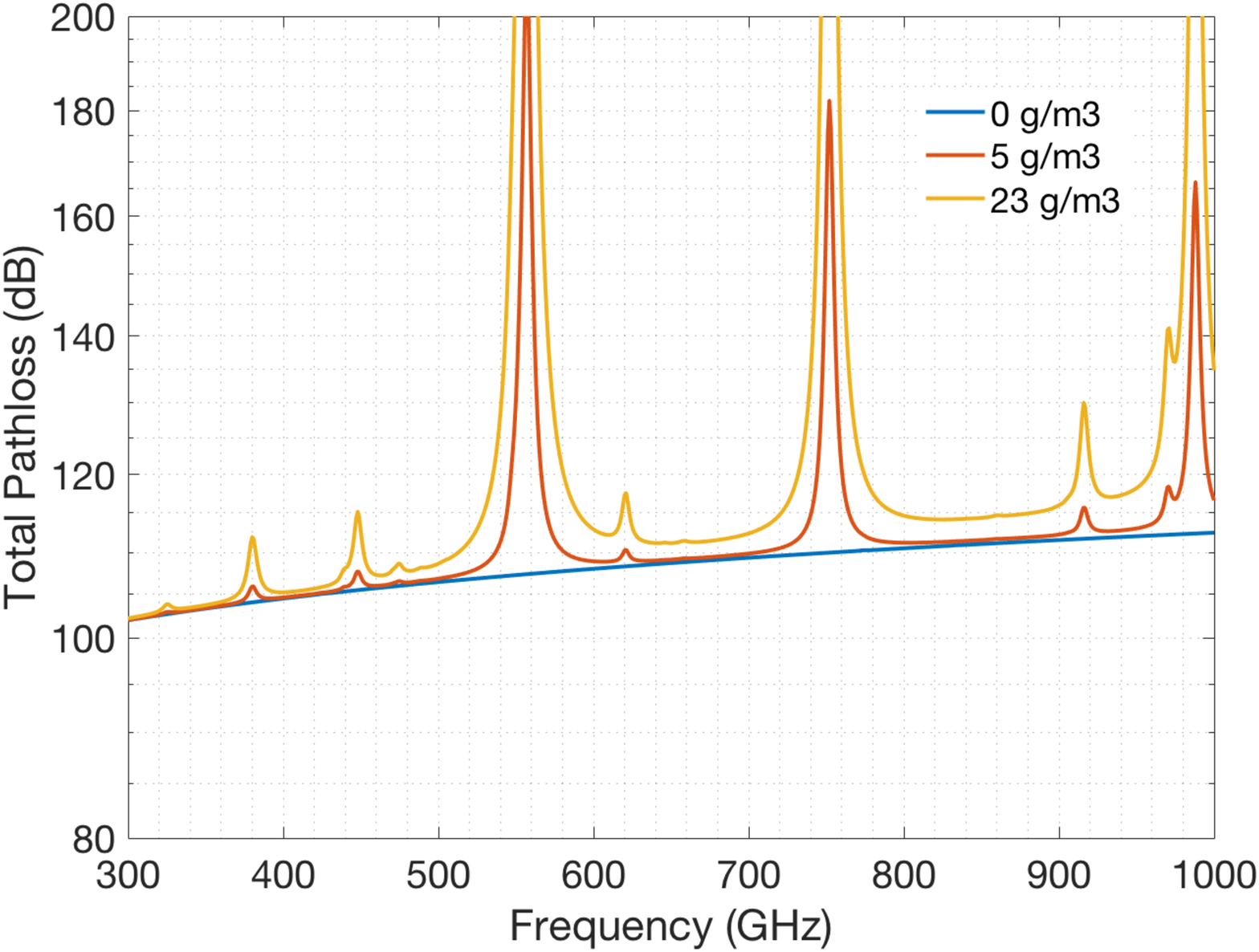}
\caption{$L_T$ at $d_j=10m$}
\label{Lt10m}
\end{subfigure}

\caption{Atmospheric loss $L_A$ \& Total Path loss  $L_T$ for different concentration of water vapor at $25 \degree C$ for $1m$ and $10m$ distance} 
\label{Loss}
\end{figure*}

One of the biggest challenges of THz band is the high path loss, a combination of high absorption loss due to water vapor concentration (shown in Fig. \ref{La1m} and \ref{La10m} \cite{ITUAtmAte}) and high spreading loss  directly proportional to the square of frequency and distance (shown in Fig. \ref{Lt1m} and \ref{Lt10m}). Thus, the total path loss is generally greater than $100 dB$, with occasional peaks approaching more than $200 dB$. Current research is focused on tackling this challenge by implementing better channel model \cite{THzChanal}, modulation schemes \cite{THzModu}, beam alignment strategy \cite{InferBeam} and antenna subarray structure \cite{HwMnAnt}. Many researchers have tried to compensate for this high path loss by increasing the transmit power or antenna gain. The current THz antennas are limited to a transmit power of $0 dBm$ and increasing it to $10 dBm$ for frequency beyond $300 GHz$ is still a challenge \cite{LiBudTHz}. Antenna gain can be improved by using more directional antenna \& narrow beamwidth \cite{HwMnAnt} to improve the received signal. Typical beamwidths range from $1 \degree$ to $20 \degree$, with a result of $47 dBi$ to $21 dBi$ gain. Although THz access-points are likely to be deployed in an indoor setting, small fluctuation for indoor devices mounted on body, hand, or head movement can still cause momentary outages, eventually resulting in lower average throughput \cite{SmallScale}.

Given that THz beams are so sensitive to user mobility, the location of AP becomes critical for efficient resource utilization. For a single user mode, the AP can be placed as close to the user equipment (UE) as possible with a direct line-of-sight (LOS). However, if we want to cover multiple UEs, the antenna has to be provided with the best degree of freedom (DoF), which is likely available only from the ceiling of the room. Nevertheless, having multiple APs will increase the chances of higher throughput and reduced outages. However, in many scenarios, it might not be possible to use these resources efficiently. Thus, it is worthwhile to consider efficient antenna resource allocation in THz bands. 

In this paper, we evaluate the optimal system and mobility parameters required for higher throughput and user coverage for both static and mobile indoor THz devices. As we move to THz, managing and monitoring parameters, such as distance, humidity, frequency, bandwidth, antenna properties, and user mobility type, becomes critical for system efficiency. Some of these parameters are either dependent on technology advancements or environmental conditions and cannot be adjusted. However, we can still adjust the antenna properties dynamically by the devices to maintain system efficiency.  For example, in THz the need for narrow beams makes it necessary that the beamwidth is selected wisely since it can lead to the ``\textit{beamwidth dilemma}"- narrower beamwidth means better gain and higher throughput, but with frequent outages due to rapid mobility and orientation of the device can eventually reduce the average throughput. We identify the system and mobility level changes required for THz communication and highlight the solutions using simulation by introducing different types of system impairments to illustrate ways of achieving the best system performance. We have also analyzed schemes through which THz resources can be used to economically and opportunistically to answer the following research questions: (i) what are the small-scale mobility-induced outage scenarios in THz, (ii) what are the mobility parameters through which we can monitor these mobility induced outages, (iii) what is the optimal antenna beamwidth for different parameters, and (iv) how many users can a single THz-AP satisfied economically in a multi-user model.

The rest of the paper is organized as follows. In Section \ref{StaticM}, we discuss the system parameter interdependencies and the tradeoffs required for a static model. In Section \ref{MobiM}, we characterize different human actions using a set of mobility parameters to identifying the mobility induced outage scenarios. In Section \ref{Eval}, we introduce environment settings to evaluate the mobility model and discuss the impact of the choice of beamwidth and antenna placement on throughput and user coverage, followed by a conclusion in Section \ref{Con}. 



\setlength\belowcaptionskip{-0.2 in}
\setlength{\abovecaptionskip}{0.2 in}
\begin{figure*}[t]
\centering
\begin{subfigure}[]{1.6 in}
\includegraphics[width=1.6 in,height=1.7 in]{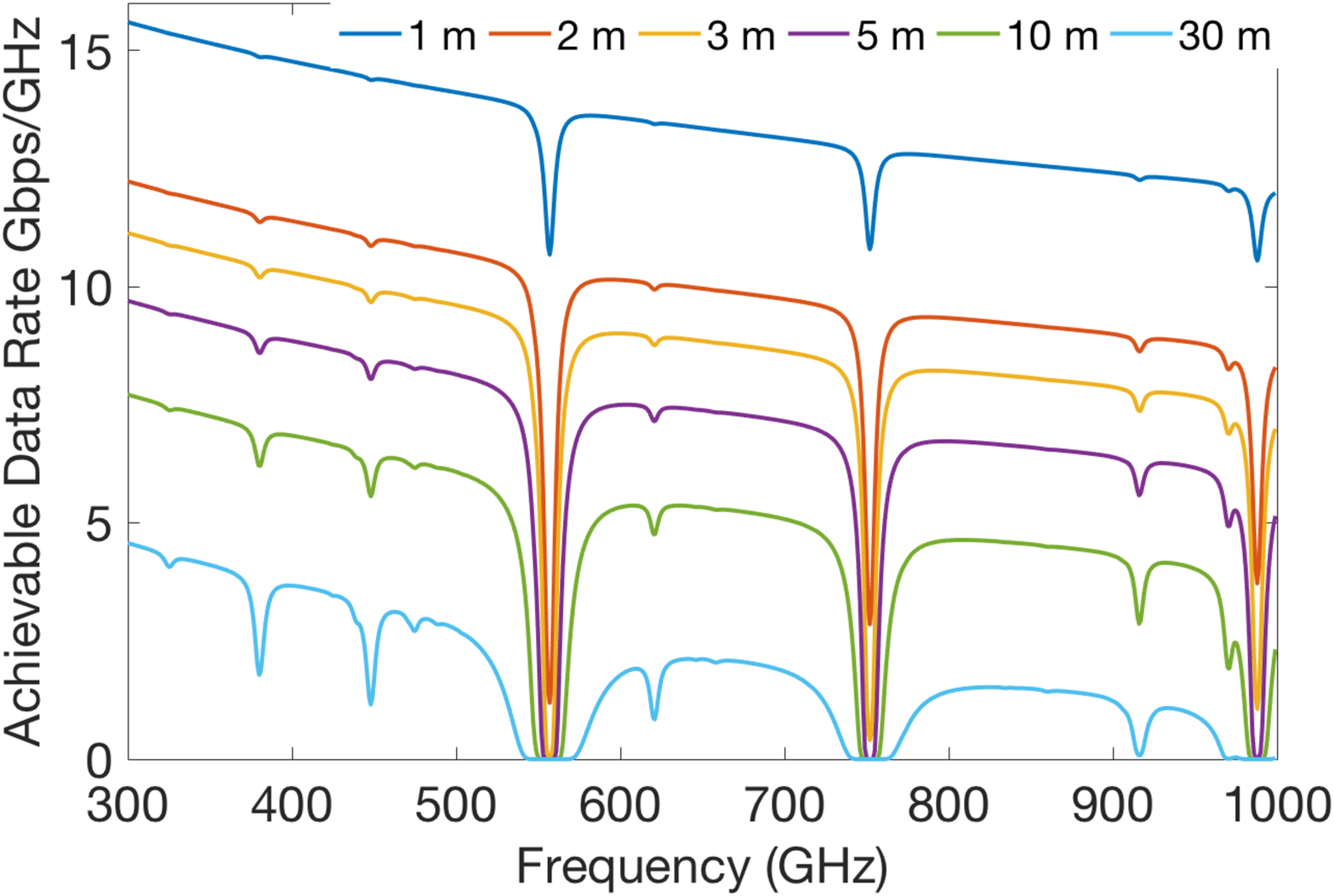}
\caption{Data Rate for $\delta =10 \degree$}
\label{AcBW10}
\end{subfigure}
~
\begin{subfigure}[]{1.6 in}
\includegraphics[width=1.6 in,height=1.7 in]{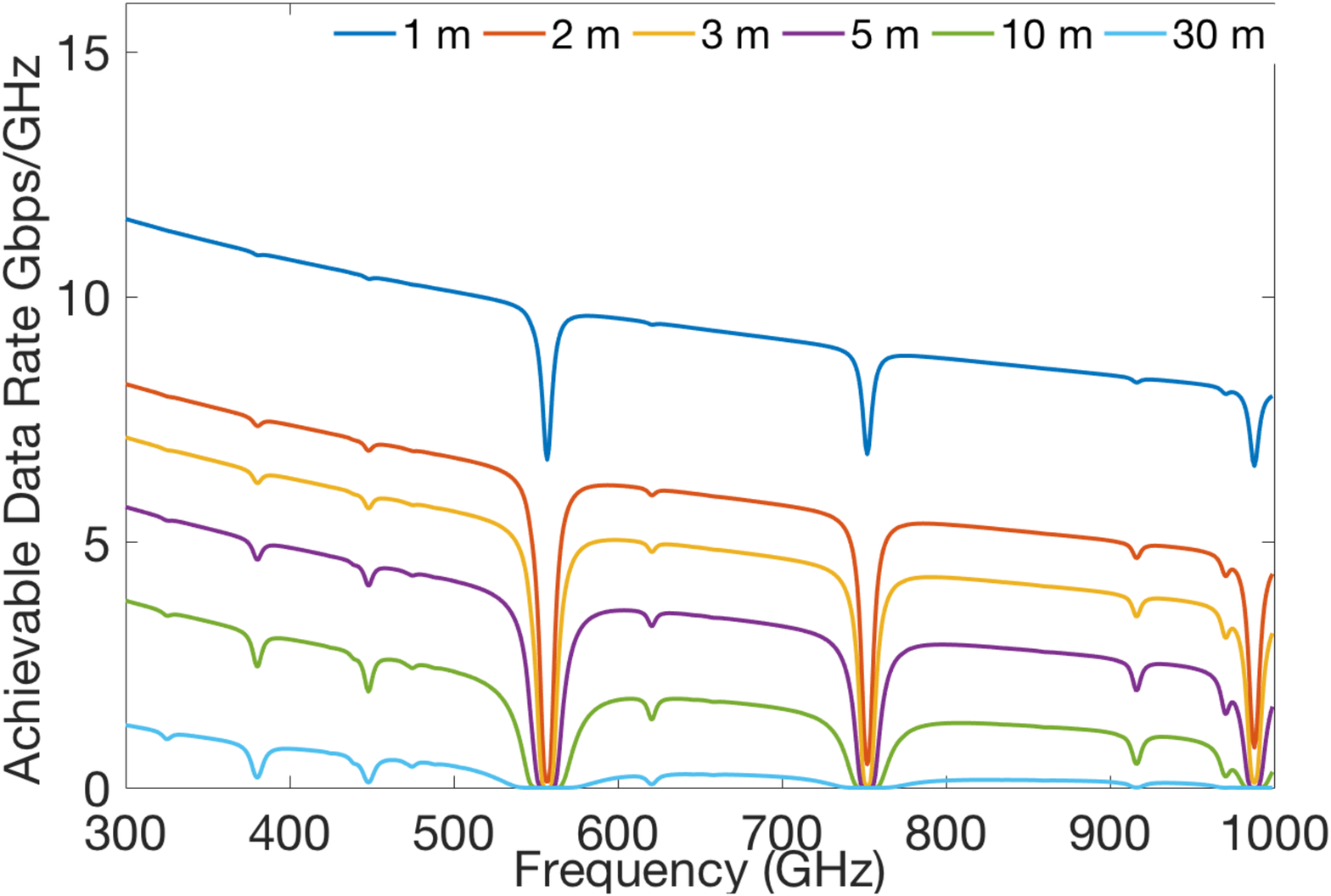}
\caption{Data Rate for $\delta =20 \degree$}
\label{AcBW20}
\end{subfigure}
~
\begin{subfigure}[]{1.6 in}
\includegraphics[width=1.6 in,height=1.7 in]{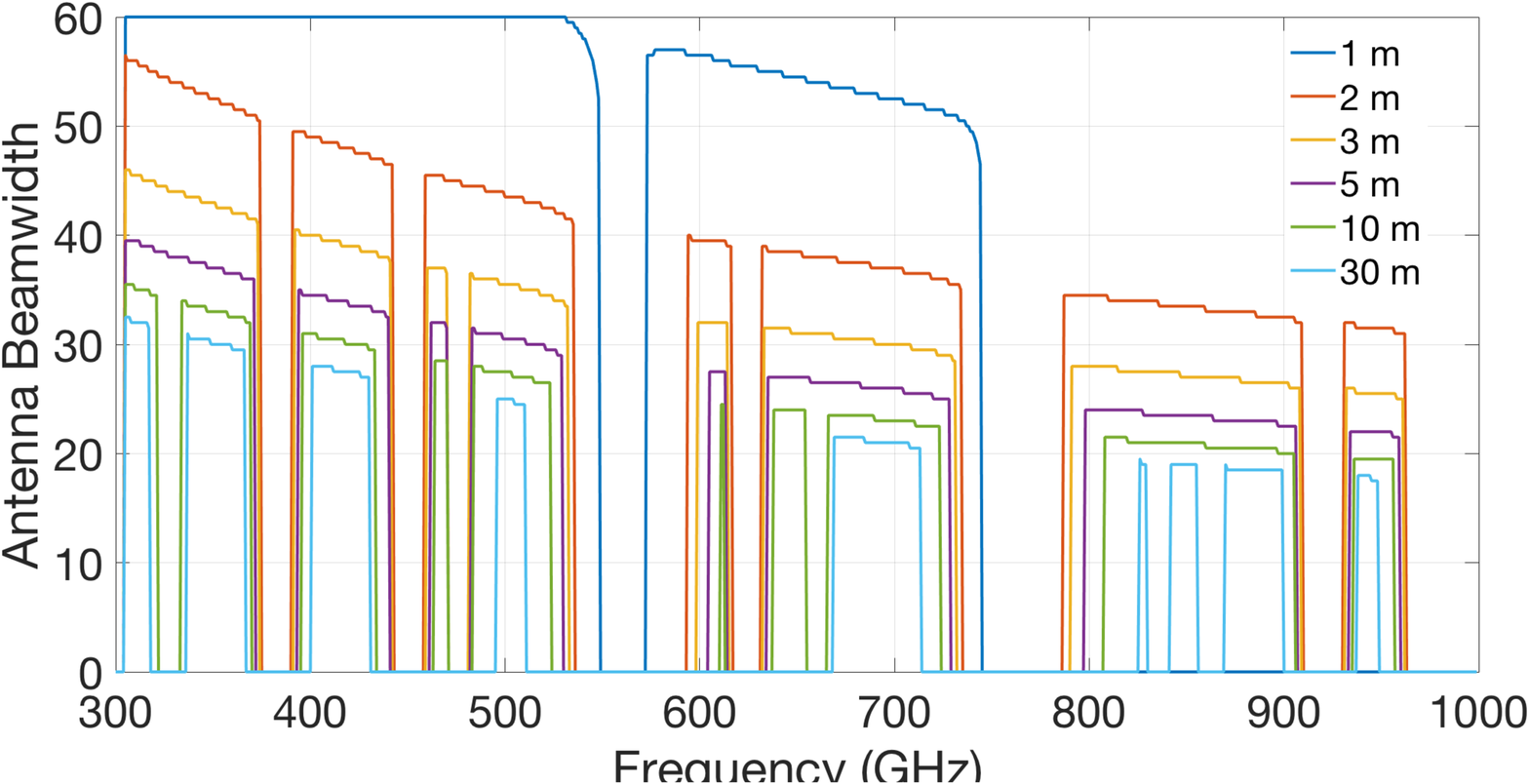}
\caption{$\delta_{min}$ for $R_j^*=10Gbps$}
\label{MR10}
\end{subfigure}
~
\begin{subfigure}[]{1.6 in}
\includegraphics[width=1.6 in,height=1.7 in]{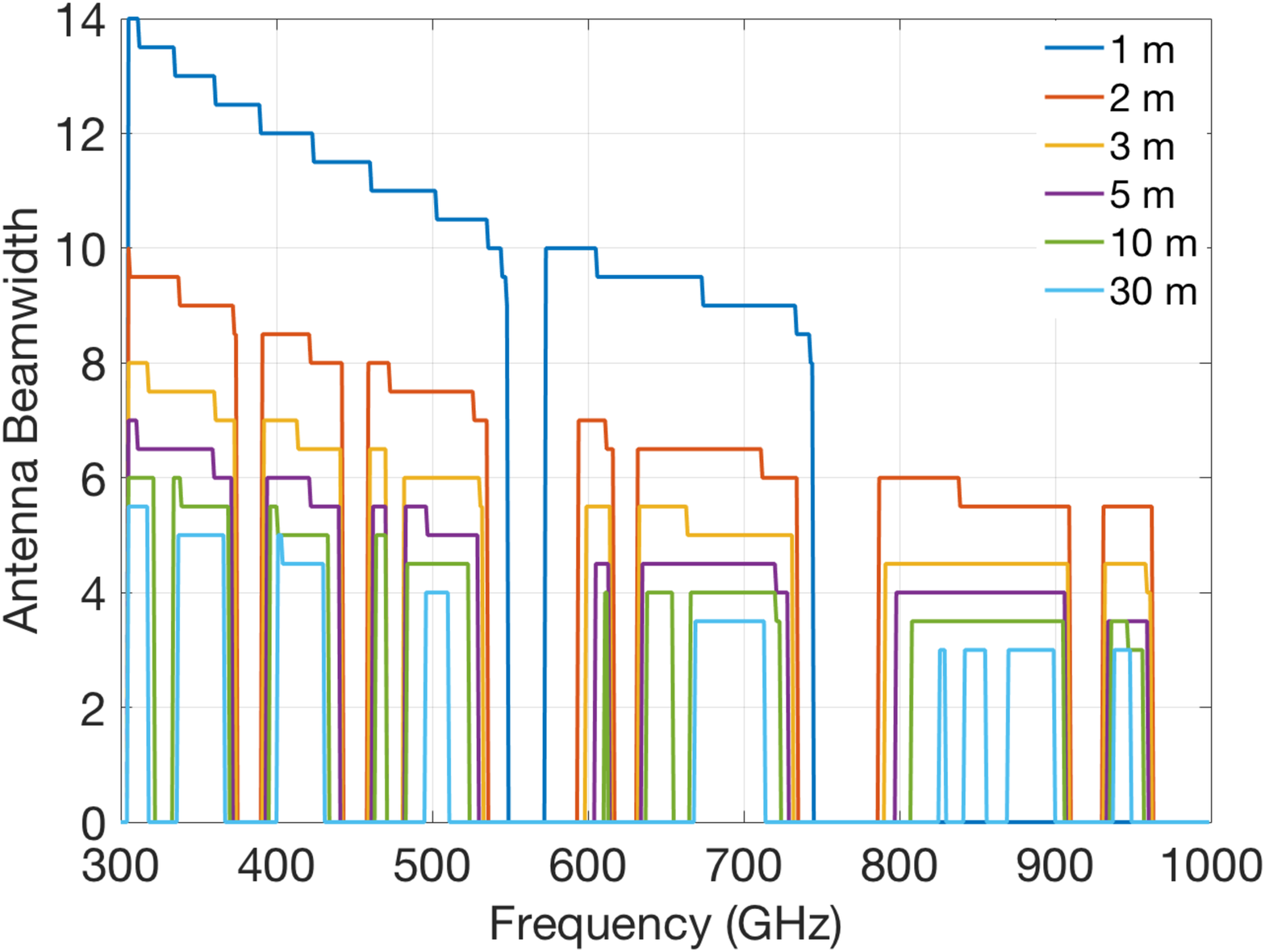}
\caption{$\delta_{min}$ for $R_j^*=100Gbps$}
\label{MR100}
\end{subfigure}

\caption{Achievable data rate and Minimum Beamwidth required for different frequency} \label{All}
\end{figure*}
    
\setlength{\intextsep}{0pt}
\setlength\belowcaptionskip{0 in}
\setlength\abovecaptionskip{0 in}
\begin{table*}[!b]  
\caption{A frequency-window lookup table for varying distances}
\centering
\begin{tabular}{ |p{1.2cm}|p{6.4cm}|p{5.7cm}|p{3.3cm}|} 
\hline
\textbf{Distance}	& 	\textbf{Average Center Frequencies $\pmb{f_c}$ in  $\pmb{GHz}$}	&	\textbf{Average Bandwidths	 $\pmb{B_j}$ in $\pmb{GHz}$ }	&  \textbf{Continuous Window Count}  \\ \hline
$\le 0.3 m$	&	424.5, 659.5, 842, 946		&	249, 173, 150, 50		& 4 ($\ge 50Ghz$), 0($< 50Ghz$) \\ \hline
$\le 1 m$		&	374.5, 498.5, 599, 684, 843, 946.5		& 149, 95, 44,120, 146, 49	& 5 ($\ge50Ghz$), 1($<50Ghz$) \\ \hline
$1-2 m$	&	340, 416.25, 497.25, 603.25, 683.25, 847, 946.5	& 80, 62.5, 88.5, 35.5, 114.5, 136, 43	& 5 ($\ge50Ghz$), 2($<50Ghz$) \\ \hline
$3-10 m$ 	&	313, 351.7, 416.8, 466.7, 504.5, 611, 681.2, 854.7, 946.2 		& 76.75, 56.75, 18.75, 56.5, 20, 104.5, 120.75, 36	& 5 ($\ge50Ghz$), 3($<50Ghz$) \\ \hline
$10-11 m$ 	& 313, 351.5, 414.75, 467, 503.5, 611.5, 646.25, 694.25, 856.75, 945.75	& 26, 45, 46.5, 16, 48, 10, 25.5, 66.5, 106.5, 29.5	& 2 ($\ge50Ghz$), 8($<50Ghz$) \\ \hline
 \end{tabular}
\label{Tab1}
\end{table*} 
    
\section{A Static Model} \label{StaticM}

Setting aside mobility, we analyze users on a snapshot basis to explore the parameter adjustments available at the system level that can be optimized to increase throughput. The analysis done in this section will not only aid in our mobility model, but also for applications that are static in nature, such as 8K TV, hologram communication, short-haul point to point communication, or even high-speed backhaul fiber replacement. 

Fig. \ref{Loss} shows that due to high path loss coupled with occasional attenuation peaks there are optimal frequency windows for THz devices to operate. These operating frequencies are mostly dependent on the distance $d_j$ for a user $j$ and water vapor concentration $\rho$. However, these factors are uncertain and dependent on the environment. Therefore, we build a static model that can adaptively select frequency windows based on distance and humidity. To do so we first need to quantify the relationship between frequency and other system parameters. 

To compensate for the huge path-losses, using more directional antennas with narrow beams can help improve the signal strength. While directional THz antennae have irregular main lobes and side lobes, for simplicity, in this paper we assume an antenna pattern with perfect conical shaped main lobe (and we recognize the implication of this assumption). For a perfect main lobe, the antenna gain $G$ can be represented as $G=\frac{X}{\delta_h \delta_v}$, where $\delta_h$ and $\delta_v$ are horizontal and vertical beamwidths in degrees, and $X$ is an aperture dependent factor, which is equal to $52,525$ for a uniformly illuminated circular aperture \cite{GainApar} and $41,253$ for uniformly illuminated rectangular aperture \cite{GainApar}. We assume that the vertical and the horizontal beams are equal, i.e., $\delta_h=\delta_v=\delta$; thus, the new antenna gain is $G=\frac{X}{\delta^2}$, which is similar to the one proposed in \cite{SmallScale}. 

Based on the antenna gain, we can calculate the achievable capacity $R_j$ at specific center frequencies $f_c$. We can calculate $R_j$ using Equation \ref{REq} \cite{MyOld}, where $B_j$ is the bandwidth, $P_t$ is the transmit power, $G_t$ and $G_r$ are gains at transmitter and receiver respectively, $L_T (f_c,d_j,\rho)$ is the total path loss and $N_o$ the noise power spectral density in $dB/Hz$. \footnote{ We assume a noise power of $-193.85 dB/Hz$, which can be calculated as $N_o=10log_{10}(N_fKT_k)$ , where $N_f$ is the noise figure of $10 dB$, K is the Boltzmann constant of $1.3810e^{-23}$, and  $T_k$ is the temperature in Kelvin.}.

\abovedisplayskip=-4pt
\belowdisplayskip=4pt
\begin{eqnarray}
R_j=B_jlog(1+\frac{P_t*G_t(\delta)*G_r(\delta)}{L_T (f_c,d_j,\rho)*N_o})
\label{REq}
\end{eqnarray}

Based on Equation \ref{REq}, we observe that the achievable data rate is mostly dependent on bandwidth $B_j$, center frequency $f_c$, separation $d_j$ and beamwidth $\delta$. A comparison of these parameters are shown in Fig.s \ref{AcBW10} and \ref{AcBW20}. It seems that the beamwidth only scales up the data rate and is not affecting the huge data rate dips. These sudden data rate dips become prominent as we increase the separation $d_j$. Thus, for higher Eucladian distance $d_j$, the continuous frequency bands decrease, resulting in decrease in additional bandwidth.  

To obtain higher throughput, it is necessary that we have continuous ``\textit{frequency-windows}" with wider bandwidths. These discontinuities on the frequency domain can be estimated using $d_j$ for a particular user and clipping of the frequency bands with huge data rate dips. For a constant $\rho$ of $5 g/m^3$ at room temperature of $25 \degree C$, we assume that an AP can either adaptively select a frequency-window or use a pre-calculated lookup table for optimal throughput. A sample frequency-window lookup table for varying distances is shown in Table \ref{Tab1}. As we keep on increasing the distnace $d_j$, the number of windows increase and the window size (or bandwidth) decreases. It is interesting to observe that at a distance of $30 cm$ we can almost avail an uninterrupted bandwidth $250 GHz$, and at $10m$ we can hardly get $50GHz$ of bandwidth. For the rest of the paper, we assume that each user can select the best frequency-window adaptively based on the environmental parameters.

Based on the analysis shown in Table \ref{Tab1}, we can estimate a relationship between the minimum beamwidth and center frequency required for a static user. For a given distance and water concentration, a theoretical bound for the minimum antenna beamwidths $\delta_{min}$ required by a center frequency $f_c$ and bandwidth $B_j$ is shown in Equation \ref{REq}, which is specific to a particular antenna aperture $X$ as explained in earlier. After clipping off the data rate dips and using Equation \ref{REq}, we can have a plot for the minimum beamwidth required for particular $f_c$ and $d_j$, as shown in Fig. \ref{MR10} and \ref{MR100}, for $B_j=10GHz$ and $X=52525$. As the requested data rate increases from $10 Gbps$ to $100 Gbps$ the $\delta_{min}$ values decreases rapidly. 

\abovedisplayskip=-4pt
\belowdisplayskip=4pt
\begin{eqnarray}
\delta_{min}=[\frac{X^2 * P_t}{L_T (f_c,d_j,\rho)*N_o*(2^{R_j^*/B_j}-1)}]^{1/4}
\label{REq}
\end{eqnarray}

With the results shown in Fig. \ref{All} we can infer that for a static model the $\delta_{min}$ values are highly dependent on the $f_c$, $R_j$ and $d_j$ values. In a static model, APs can be strategically placed to obtain higher throughput based on the frequency allocation available through regulators and environment impairments. For example, even with a small frequency window of $10 GHz$ and a high Euclidian distance of $30m$, one can still obtain $10 Gbps$ speed at a bare minimum beamwidth of $20 \degree$.



\setlength\belowcaptionskip{-0.2 in}
\setlength{\abovecaptionskip}{0 in}
\begin{figure*}[t]
\centering
\includegraphics[width=6 in,height=3.1 in]{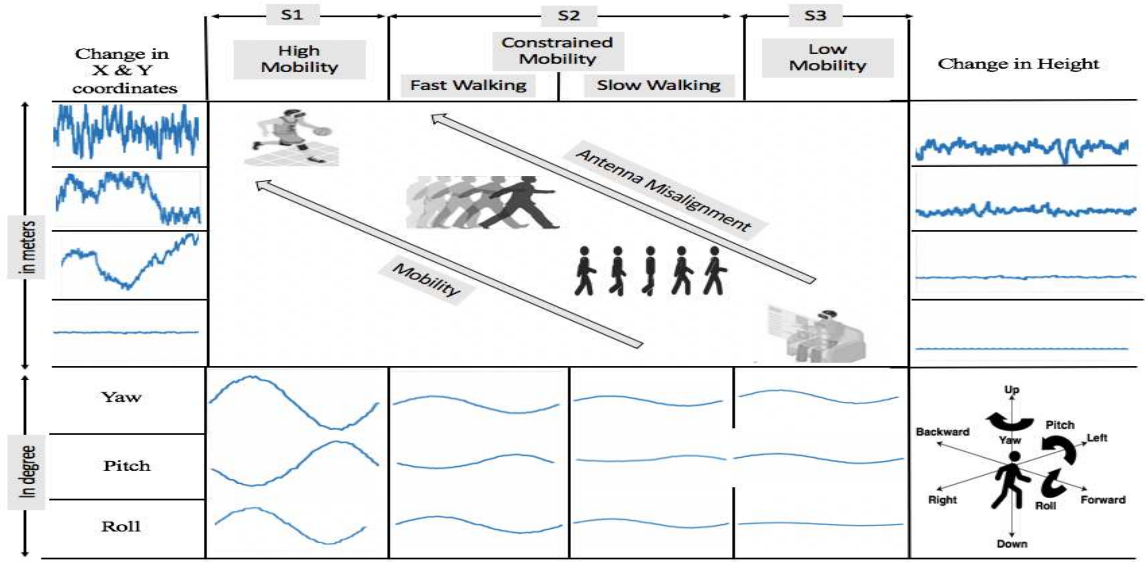}
\caption{Indoor Mobility Model  showing the patterns associated with the Six-DoF of Human Body for different Action Types}
\label{MobSum}
\end{figure*}

\section{A Mobility Model } \label{MobiM}

The analysis shown in Section 2 is true for static users and can be used to a certain extent by mobile users on a snapshot basis to select optimal frequency windows. However, when we add the impact of outages due to mobility of the users and orientation of the devices, the time-average throughput changes and so does the beamwidth values. The bare minimum beamwidth (shown in Fig. \ref{MR10} and \ref{MR100}) is not enough to maintain a time-average throughput as requested by the user. Before we estimate these optimal values, we need to identify these indoor small-scale mobility types.  

In the future, indoor mobile THz devices will not only be limited to handheld displays, or Head Mounted Device (HMD) \cite{HMDAR}, but also augmented reality contact lenses, body area network (BAN), and nanobots. For lower beamwidths we need almost perfect alignment between the TX-RX antennas; thus, making these devices highly sensitive to fluctuations of hand, body, head (maybe even eye) movements. One of the highest throughput demanding use case for THz is VR \& AR Games, which requires fast body and hand movements. Resulting in a user moving in all Six Degree of Freedom (DoF) as shown in the bottom right corner of Fig. \ref{MobSum}. These games consist of bursts of high-intensity movements followed by slow movements, which can result in frequent misalignments. Although these VR games are sparsely used today, other applications such as VR chairs and VR goggles are gaining popularity among users to watch movies or play games. Compared to the action games, the body movements are relatively reduced in this scenario, yet these devices are still susceptible to misalignments due to yaw, pitch and roll motions of the body. A more constrained mobility scenario with low body fluctuations will be downloading or streaming uncompressed video while walking or standing in front of a kiosk or shared hotspot. 

Whether it is a high-intensity game or even a constrained indoor mobility scenario, such as walking and standing, THz communication links, with narrow beamwidths, are susceptible to frequent misalignments. However, the best part of this movement is that they follow patterns. For example, in the case of action games, there are particular movements which are game specific and are repeated quite often, or in case of normal walking, the body repeats a particular velocity and movement pattern. In \cite{Bobbing} it was shown that normal human walking is subjected to \textit{bobbing}, i.e., the up and down movement of body\textquotesingle s center of mass. The rate of bobbing is directly proportional to the velocity of the human. Thus, fast pace walking is more prone to misalignments than slow pace walking. Furthermore, while walking the yaw, pitch and roll motions of the body emulate a sinusoidal wave \cite{YawSin}. Even when a human is standing or in sedentary, it tends to move in small amounts in all the Six-DoF. The change in position while sedentary might look small but in scenarios with beamwidths less than (nearly) $5 \degree$ can result in major misalignments. These features of human body movement can be parameterized for specific service types to provide the best QoS. Thus, there is a need to model these indoor mobility patterns using a set of parameters that can keep track of the Six-DoF. We name this parameter $\Theta=[\Delta x, \Delta y, \Delta z, \Delta \omega_y, \Delta \omega_p, \Delta \omega_r]$ as the \textit{system instability parameter}, where the rate of change in $x$, $y$ and $z$ coordinates (\textit{location parameters}) are represented by $\Delta x, \Delta y, \Delta z$ and change in yaw, pitch, and roll (\textit{rotation parameters}) are represented by $\Delta \omega_y, \Delta \omega_p, \Delta \omega_r$. 

Given the myriad number of multiple and small positional changes in the six dimensions of body movement, it is necessary that we have a model that imitates the human body based on service type. We bucket these activities into three major service type: (a) High Mobility (e.g., high intense games), (b) Constrained Mobility (e.g., fast and slow walking), and (c) Low Mobility (e.g., sitting or standing). A summary of the service types and the pattern of $\Theta$ values are shown in Fig. \ref{MobSum}. Each service type is unique in both the mobility pattern and the demand for QoS. For service type S1 (High Mobility) there is random fluctuation in all six dimensions of the body. For service type S2 (Constrained Mobility) there is constant change in all six dimensions and is mostly proportional to the velocity of the body. We further divide the S2 into fast and slow walking. Since both fast and slow walking can cause bobbing, affecting the $\Delta z$ and the rotation parameters, it is necessary that we monitor both of these cases separately in a constrained mobility scenario. For service type S3 (Low Mobility), it emulates more of a sedentary human, with very little change in the location parameters; yet, incorporates periodic oscillations in the rotational parameters. Detailed analysis shows that each of these parameters behaves in a particular pattern for a given type of movement. For example, research  \cite{YawSin} on movement patterns of human walking showed sinusoidal yaw of the body with a peak amplitude of $4 \degree$. Moreover, the roll and pitch rotation also represented a sinusoidal wave, where the pitch amplitude decelerated and accelerated alternatively for each left and right roll movement of the body, as shown in Fig. \ref{MobSum}. This shows that beamwidths narrower than body rotation may suffer from frequent misalignments if not monitored properly.



\section{Optimal Beamwidth for Mobile Users} \label{Eval}

For the simulations conducted in the rest of the paper, we assume a mobility that emulates $\Theta$ patterns mentioned in Fig. \ref{MobSum} and with values shown in Table \ref{Tab2} \cite{Bobbing}\cite{YawSin}\cite{RobotExercise}\cite{HTCVive}. Although the values in Table \ref{Tab2} are mostly dependent on where the device is mounted (e.g., head, hand or eyes). For simplicity we consider the general body movement. This assumption is logically valid, since a human being\textquotesingle s hand, head and body movements are somewhat interlinked. For example, if we move our hand to the left (or right) to grab an object our torso should also yaw in that direction. 

\setlength{\intextsep}{0pt}
\setlength\belowcaptionskip{0.1 in}
\setlength\abovecaptionskip{0 in}
\begin{table}[t]  
\caption{User Mobility Parameters}
\centering
\begin{tabular}{ |p{0.1cm}|p{1.3cm}|p{1.5cm}|p{1.7cm}|p{0.5 cm}|p{0.5 cm}|p{0.2 cm}|} 
\hline
\multicolumn{2}{|c|}{\textbf{Service Types}}   & $\pmb{\Delta x}$,$\pmb{\Delta y}$ & $\pmb{\Delta z}$  & $\pmb{\omega_y}$  & $\pmb{\omega_p}$  & $\pmb{\omega_r}$ \\ \hline
\multicolumn{2}{|c|}{S1} & $1 \pm 0.5$ &	$0.5 \pm 0.05$	& $15.5 \degree$ & $13.8 \degree$ & $15 \degree$ \\ \hline
\multirow{2}{*}{S2}	& Fast Walk	& $0.9 \pm 0.7$  & $0.094 \pm 0.02$	 & $4 \degree$ & $5 \degree$ & $5 \degree$ \\ \cline{2-7}
	 			& Slow Walk & $0.3 \pm 0.27$  & $0.015 \pm 0.003$ & $ 3 \degree$ & $3 \degree$	& $3 \degree$ \\ \hline
\multicolumn{2}{|c|}{S3} & $0 \pm 0.01$	& $0 \pm 0.001$	 & $5 \degree$	& $3 \degree$	& $1 \degree$ \\ \hline
 \end{tabular}
\label{Tab2}
\vspace{-6mm}
\end{table}

\begin{figure}[t]
\centering
\includegraphics[width=3 in,height=1.7 in]{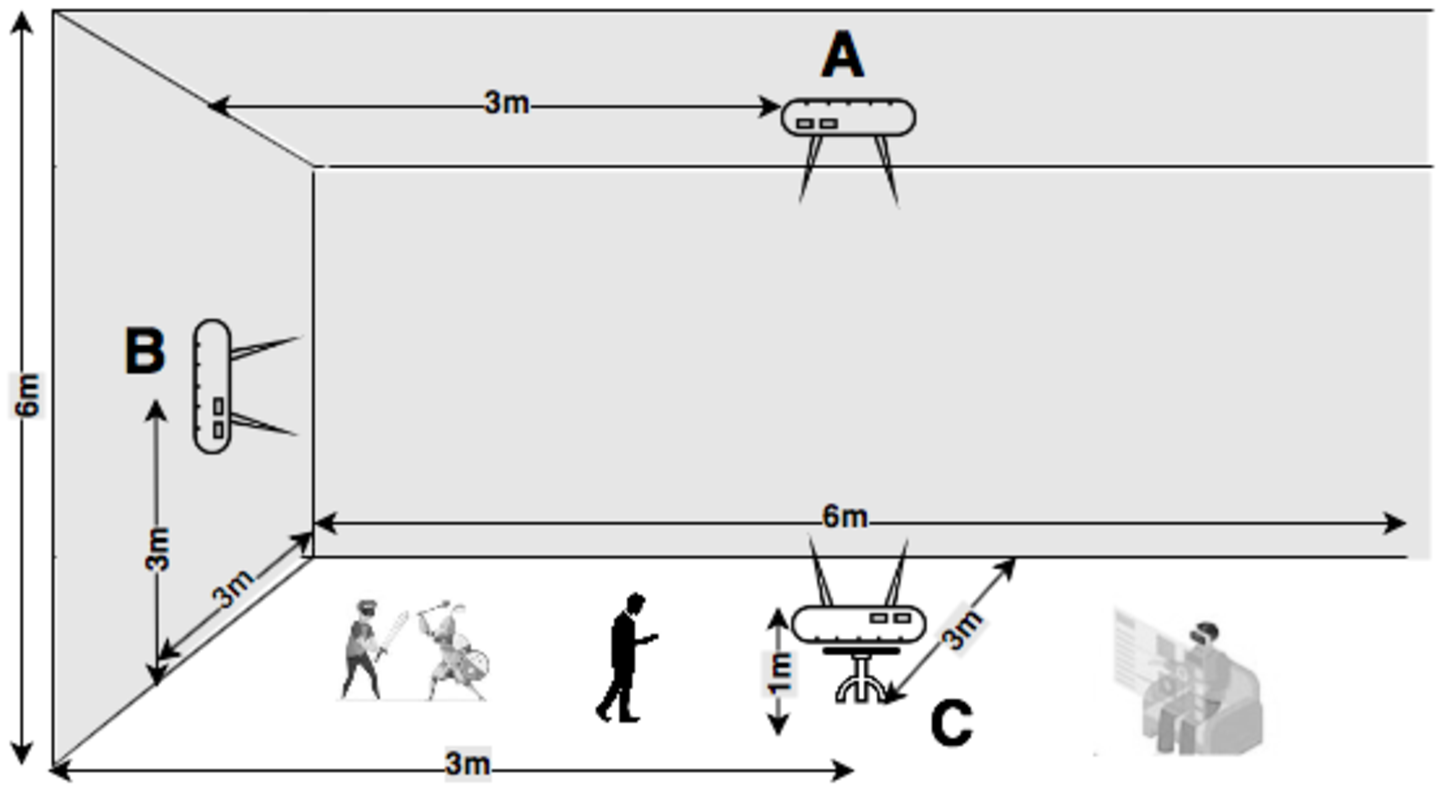}
\caption{Environment setting and different AP placement}
\label{APplace2}
\end{figure}

We consider a room shown in Fig. \ref{APplace2} with $M$ active users following a random way point model and requesting $10 Gbps$ data rate per time slot. Each user can belong to either of the three service types shown in Fig. \ref{MobSum}. Therefore, the instability parameter for a user $j$ at any instance of time is defined as $\Theta_j^t$, $t \in {S1,S2,S3}, j \in {1,2, \cdot M}$. For a fixed location of an AP ($x_a,y_a,z_a$) and a random user coordinates $\Theta_j^t$, we can define $d_j^t$ as the distance between the AP and UE $j$. To find a user\textquotesingle s location, we use a random way point model coupled with the mean and standard division of the location parameters mentioned in Table \ref{Tab2}. Please note that to calculate the resultant $z$ coordinate, we considered an average human standing height of $1.5m$ for S1 and S2, and a sitting height of $1.2m$ for S3. To calculate the rotational changes, we use the peak values of the rotation parameters mentioned in Table \ref{Tab2} to replicate sinusoidal wave patterns shown in Fig. \ref{MobSum}. Please note that while a body is in motion, specifically in S1 and S2 scenarios, the pitch and roll movements experience alternative deacceleration and acceleration, which is not present when a body is sedentary, i.e., in case of S3 \cite{YawSin}. We also allow the rotational movements through a small amount of noise to make the rotations more realistic. We assume the APs to be at three different locations, as shown in Fig. \ref{APplace2}.  Instead of a random AP placement, we explored 3 AP locations, Scenario A (middle of the ceiling), Scenario B (middle of the wall) and Scenario C (middle of the room on a $1m$ high table). As we move the AP away from one location in the room to another, the optimal beamwidth changes. Each AP location has its own advantages and disadvantages and is explained later.

\begin{figure*}[t]
\centering
\begin{subfigure}[]{2 in}
\includegraphics[width=2 in,height=1.5 in]{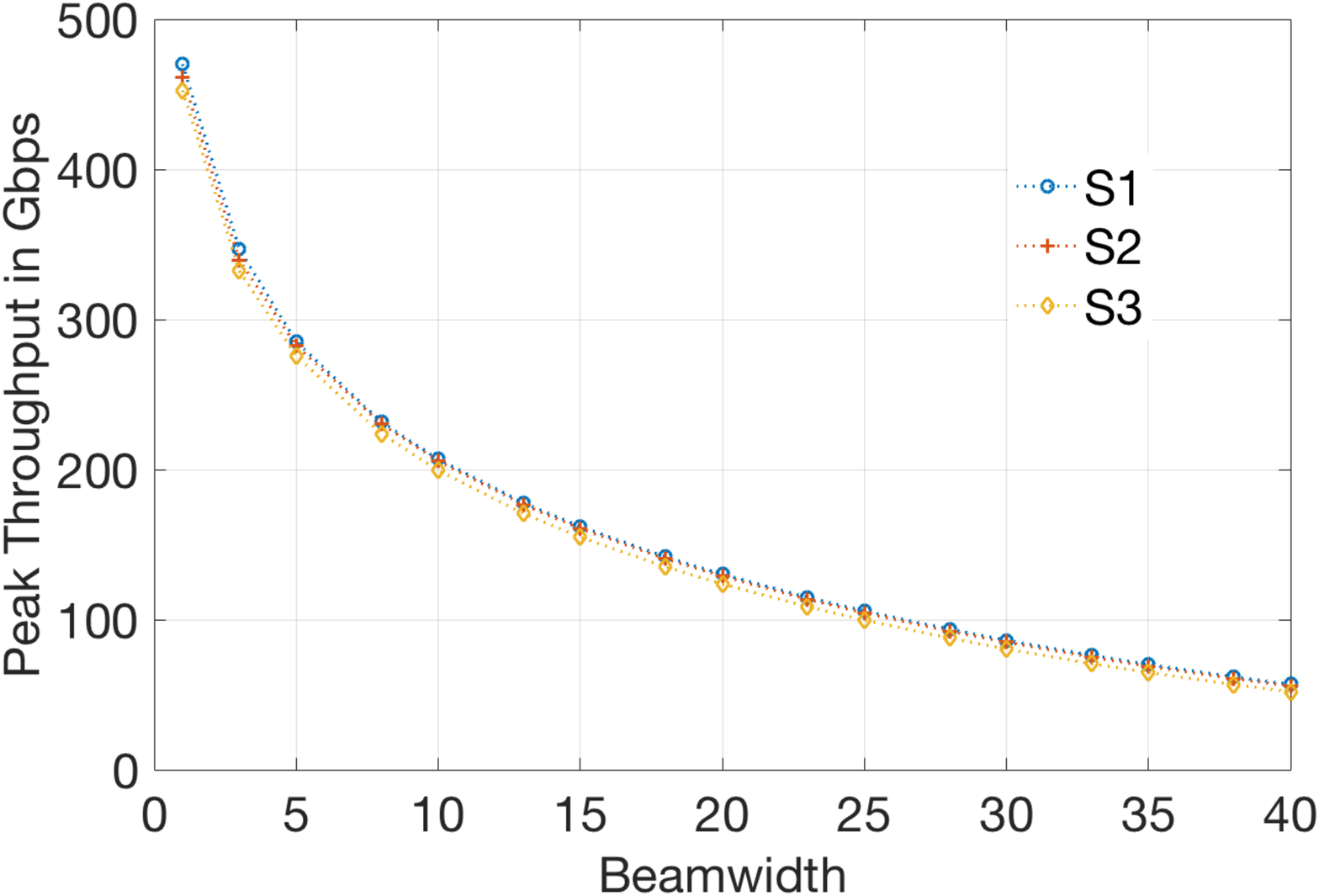}
\caption{Scenario A}
\label{TA}
\end{subfigure}
~
\begin{subfigure}[]{2 in}
\includegraphics[width=2 in,height=1.5 in]{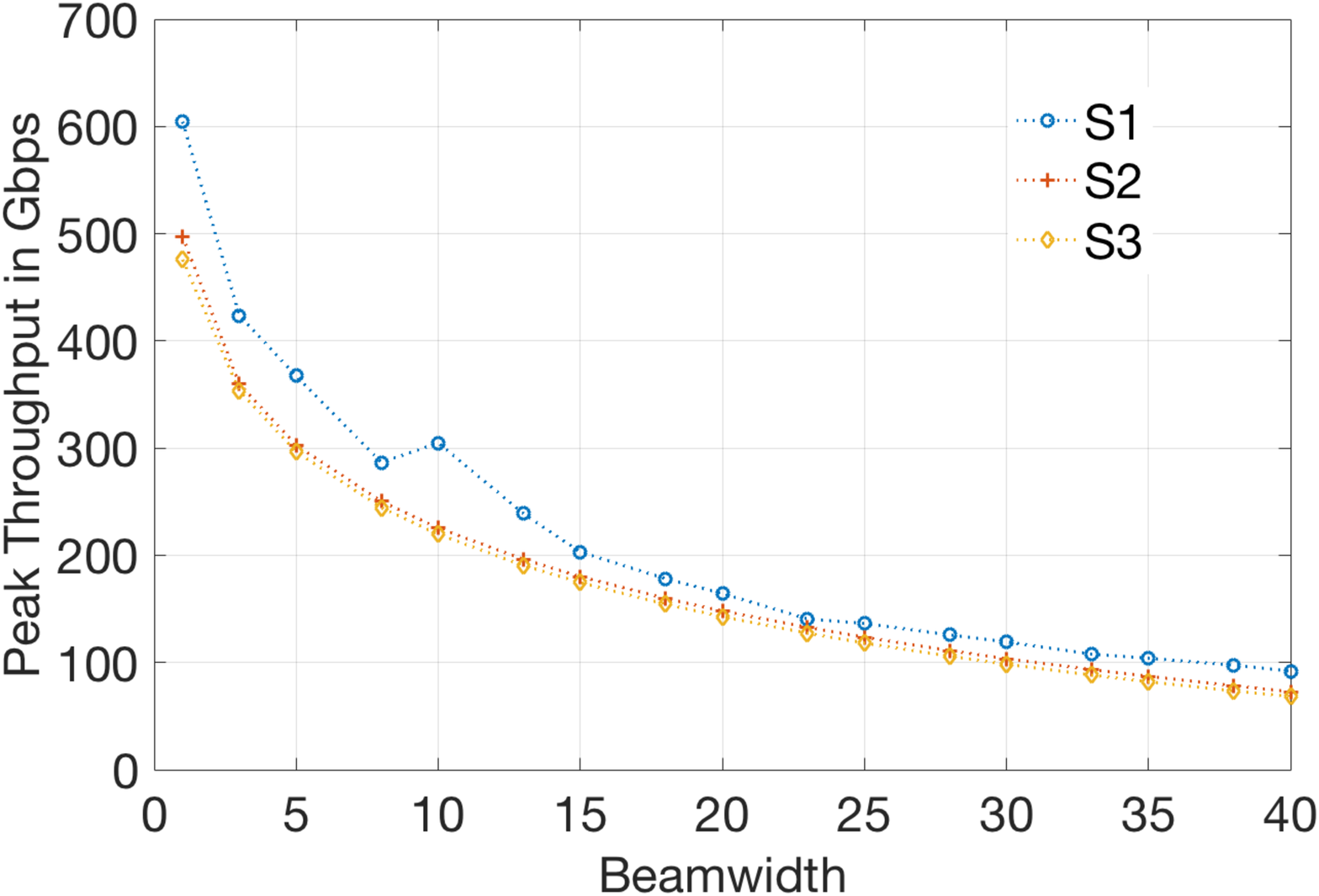}
\caption{Scenario B}
\label{TB}
\end{subfigure}
~
\begin{subfigure}[]{2 in}
\includegraphics[width=2 in,height=1.5 in]{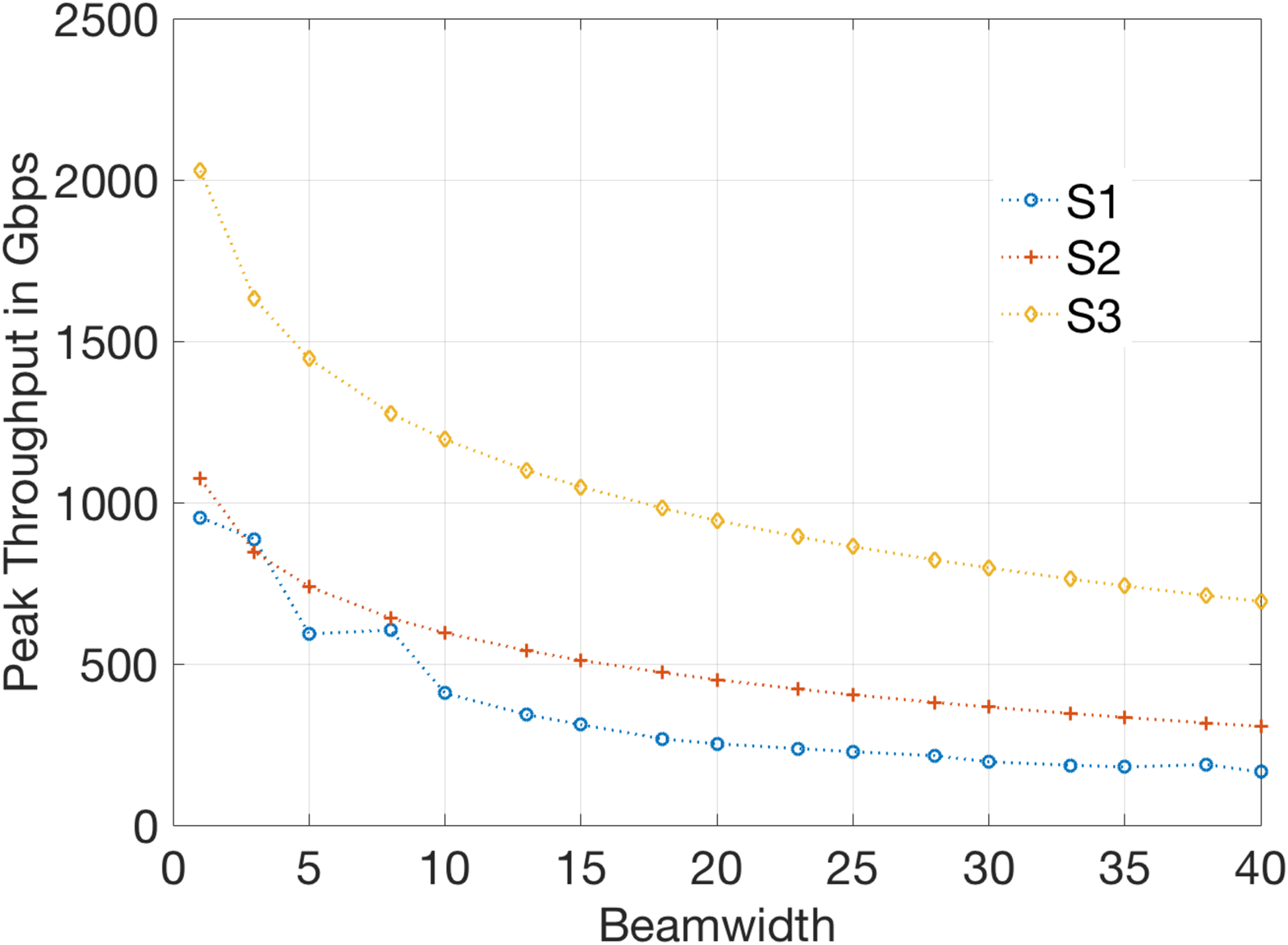}
\caption{Scenario C}
\label{TC}
\end{subfigure}
\caption{Peak Throughput for different service types and AP placements} \label{Thall}
\end{figure*}

It is obvious as the beamwidth increases the gain will decrease, eventually decreasing the throughput, as shown in Fig. \ref{Thall}. However, the AP placement also has a significant effect on the throughput. As we move from Scenario A to C, the throughput values increase rapidly, irrespective of the service types. Although Scenario A has a better coverage compared to Scenarios B and C, reducing the height of the antenna critically effects the system efficiency. Therefore, it will be beneficial to place the APs as close to the users as possible (due to high sensitivity to distance). For scenario A, all three service types have overlapping throughput values, and the gaps widen as we move to Scenarios B and C. Scenario B and C are effectively able to capture the yaw and pitch movements of the devices; thus, resulting in lower outages. 

\setlength\belowcaptionskip{-0.3 in}
\setlength{\abovecaptionskip}{0 in}

\begin{figure*}[t]
\centering
\begin{subfigure}[]{2 in}
\includegraphics[width=2 in,height=1.5 in]{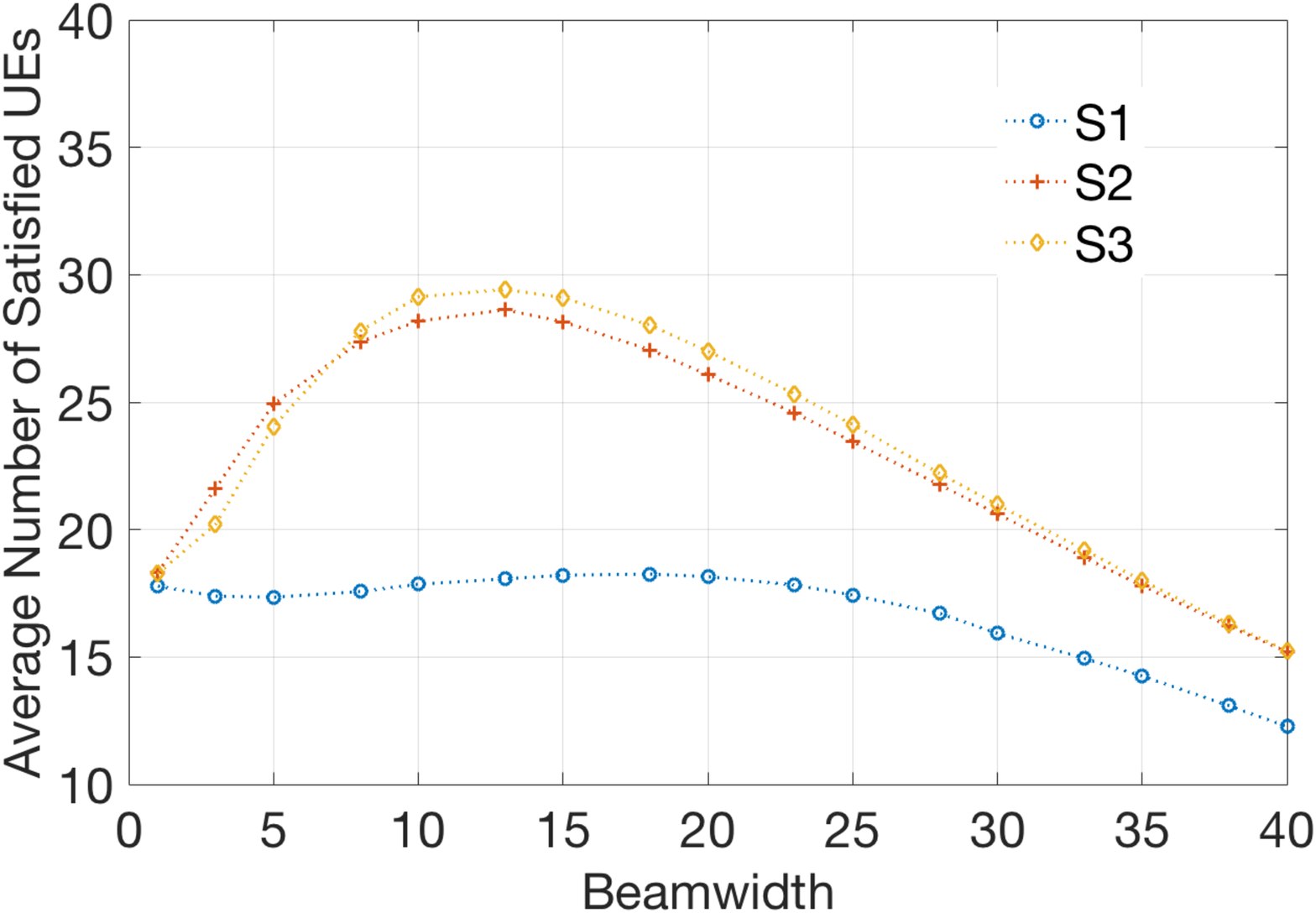}
\caption{Scenario A}
\label{UA}
\end{subfigure}
~
\begin{subfigure}[]{2 in}
\includegraphics[width=2 in,height=1.5 in]{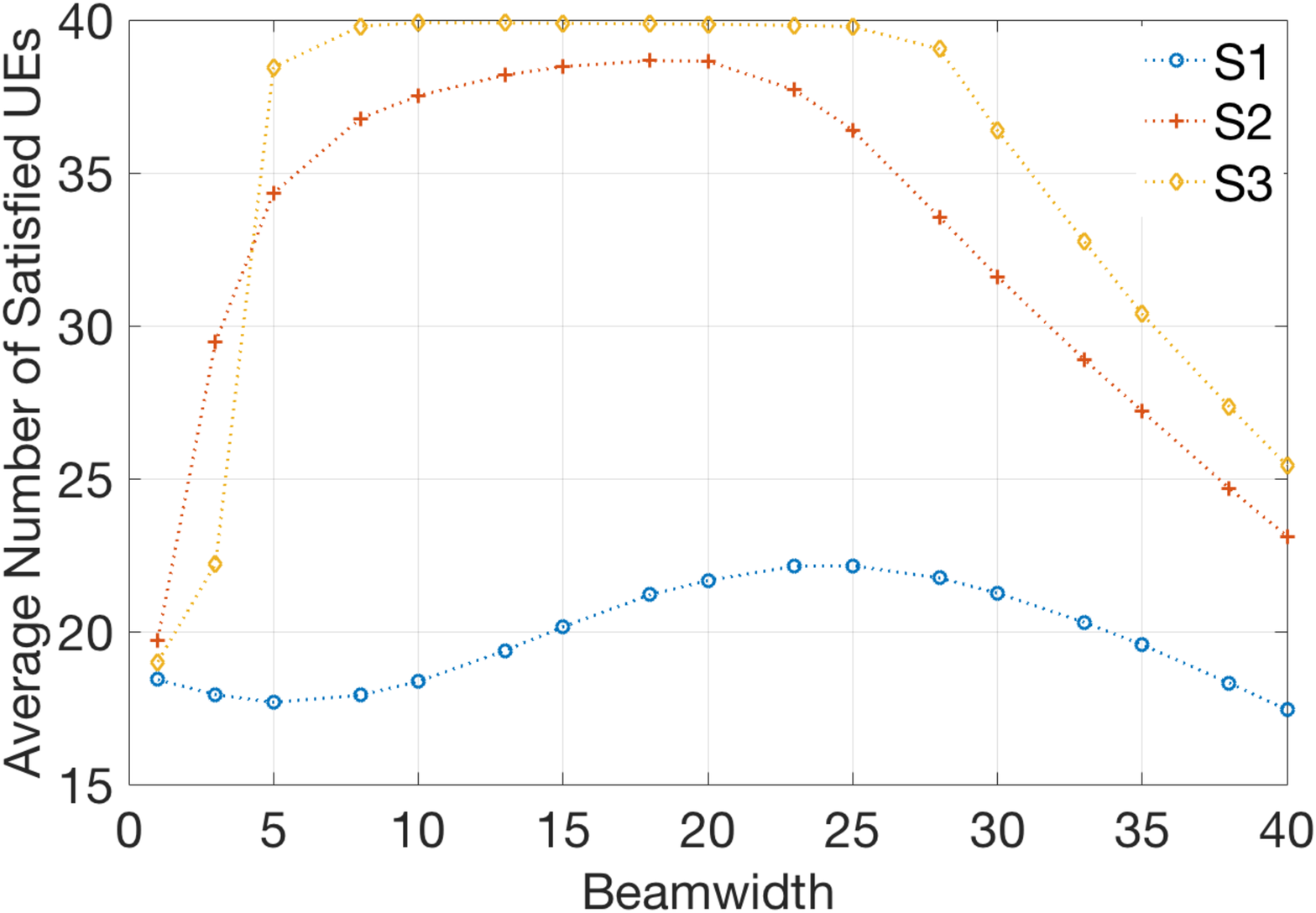}
\caption{Scenario B}
\label{UB}
\end{subfigure}
~
\begin{subfigure}[]{2 in}
\includegraphics[width=2 in,height=1.5 in]{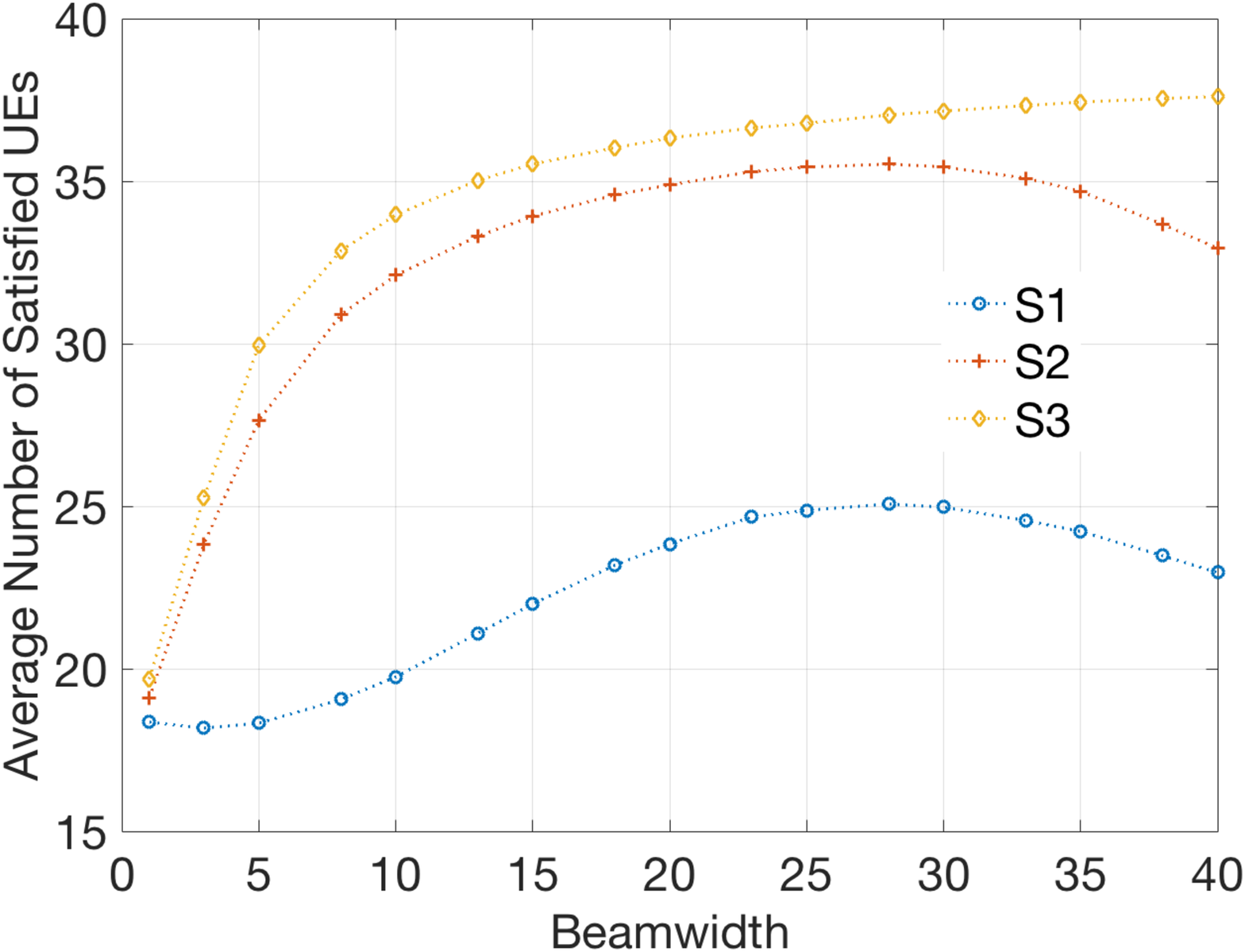}
\caption{Scenario C}
\label{UC}
\end{subfigure}

\caption{Average User Coverage for different service types and AP placements} \label{UEall}
\end{figure*}

Although getting higher throughput is necessary, for a multi-user scenario the critical system objective is to have a higher user coverage. For service type S2 and S3, Scenario B type AP placement shows on an average higher coverage compared to Scenario A and C, as shown in Fig. \ref{UEall}. While for Service type S1, on an average Scenario C shows higher coverage. We can even observe optimal beamwidth values for the different service types and AP placements. The nature of the curves is subjected to the human body movement as summarized in Fig. \ref{MobSum}. If the rotation angle of the body (yaw, pitch, and roll) is higher than the beamwidth of the antenna, there are likely to be higher outages, resulting in a lower average user count. It is quite evident for S1, where the rotation angles are high and so are the outages are higher for very narrow beamwidths. On the contrary,  S2 and S3 have relatively lower rotation angles, resulting in very low outages for lower beamwidth values. 

The numerical evaluations are shown in Fig. \ref{Thall} and \ref{UEall} imply that for some fixed system parameters (frequency windows) there exists a $\delta_{opt}$ value that can help solve the ``\textit{beamwidth dilemma}" issue. There is a tradeoff between throughput and user-coverage based on: (a) device mobility parameter $\Theta^t, t \in {S1,S2,S3}$, and (b) AP placement $x_a,y_a,z_a$. For example, to achieve higher user coverage for service type S1 users the $\delta_{opt}=23\degree$ for Scenario B, and $\delta_{opt}=25 \degree$ for Scenario C. It seems for service type S2 and S3 type users the $\delta_{opt}$ values are lower, since with constrained and low mobility it is easier to satisfy multiple users at lower beamwidth values with minimum outages. It is harder to get higher coverage for S1 type users, due to the high fluctuation. 

Although there are pros and cons to different AP placement type, each placement type can be used for different applications. Scenario A is likely to be in applications for VR games, office use or even in public safety events, where a drone (cluster head), out of a swarm of drones, can emulate a THz AP to get the maximum user coverage. On the other hand, Placement type B and C are more likely for applications for indoor users, ultra-high-speed kiosks or used as personal hotspots at waiting lounges. The analysis supports the need for adaptively changing beamwidths (and other system parameters) based on service type and antenna placement. 



\setlength\belowcaptionskip{-0.2 in}
\setlength{\abovecaptionskip}{0.2 in}

\setlength\belowcaptionskip{-0.2 in}
\setlength{\abovecaptionskip}{0.2 in}

\section{Conclusion} \label{Con}

In this paper, we analyze parameters, such as frequency, beamwidth, AP placement, and mobility type that are necessary for higher throughput and user coverage for indoor THz communications. To use THz resource more economically and opportunistically, we (a) perform numerical analysis for a single AP, (b) characterize the small-scale mobility type based on the human actions, and (c) quantify the relationship between beamwidth and other system parameters. We observe that for static applications a theoretical bound for the beamwidth can be drawn based on the system parameters such as frequency and requested data rate. While for the mobile applications, we observe that the services type and the nature of the device orientation can significantly affect the optimal beamwidth. In a mobile scenario, the optimal beamwidths are not only dependent on service type, but also the AP location. Our parameter analysis for static and mobile applications will help address issues related to THz deployment (frequency and AP placement) and ``\textit{beamwidth dilemma}" for indoor settings. Systems can dynamically change the $\delta_{opt}$ values based on four critical factors: (a) objective functions (throughput or user coverage), (b) frequency windows, (c) mobility parameters $\Theta^t$, and (d) AP locations.

\end{document}